\documentclass{elsart}

\usepackage{graphicx}

\usepackage{amssymb}
\usepackage{amsmath}

\DeclareMathOperator{\spur}{sp}
\DeclareMathOperator{\Spur}{Sp}
\DeclareMathOperator{\re}{Re}

\begin{document}

\begin{frontmatter}




\title{Universal properties of frustrated spin systems: $1/N$-expansion and renormalization group approaches}


\author[IMP]{A.N. Ignatenko}
\ead{ignatenko@imp.uran.ru}
\author[IMP]{V.Yu. Irkhin}
\author[IMP,MP]{A.A. Katanin}

\address[IMP]{Institute of Metal Physics, Kovalevskaya Str., 18, 620041 Ekaterinburg, Russia}
\address[MP]{Max-Planck-Institut f\"ur Festk\"orperforschung, 70569 Stuttgart, Germany}

\begin{abstract}
We consider a quantum two-dimensional $O(N)\otimes O(2)/O(N-2)\otimes O(2)_{\text{diag}}$ nonlinear sigma model for frustrated spin systems and formulate its $1/N$-expansion which involves fluctuating scalar and vector fields describing kinematic and dynamic interactions, respectively. The ground state phase diagram of this model is obtained within the $1/N$-expansion and $2+\varepsilon$ renormalization group approaches. The temperature dependence of correlation length in the renormalized classical and quantum critical regimes is discussed. In the region $\rho_{\text{in}}<\rho_{\text{out}}$, $\chi_{\text{in}}<\chi_{\text{out}}$ of the symmetry broken ground state ($\rho_{\text{in,out}}$ and $\chi_{\text{in,out}}$ are the in- and out-of-plane spin stiffnesses and susceptibilities), where the mass $M_{\mu}$ of the vector field can be arbitrarily small, physical properties at finite temperatures are universal functions of $\rho_{\text{in},\text{out}}$, $\chi_{\text{in},\text{out}}$, and temperature $T$. For small $M_{\mu}$ these properties show a crossover from low- to high temperature regime at $T\sim M_{\mu}$. For $\rho_{\text{in}}>\rho_{\text{out}}$ or $\chi_{\text{in}}>\chi_{\text{out}}$
finite-temperature properties are universal functions only at sufficiently large $M_{\mu}$. The high-energy behaviour in the latter regime is similar to the Landau-pole dependence of the physical charge $e$ on the momentum scale in quantum electrodynamics, with mass $M_{\mu}$ playing a role of $e^{-1}$. The application  of the results obtained to the triangular-lattice Heisenberg antiferromagnet is considered.

\end{abstract}

\end{frontmatter}

\section{Introduction}

The description of frustrated systems is a long-discussed subject \cite{A73,CSS94,CCL90}. These systems pose an important and interesting problem of condensed matter physics, and also provide a test of various methods of quantum field theory \cite{K88,A93,NPRG04}. Competing interactions leading to frustration favor strong quantum fluctuations and therefore tune the system towards quantum phase transition (QPT) into the phase without spontaneous symmetry breaking.

Examples of frustrated systems are the Heisenberg triangular lattice antiferromagnets (TLAF). In particular, VCl$_{2}$ and VBr$_{2}$ are layered compounds with the triangular lattice structure within a layer. The corresponding Heisenberg model with nearest-neighbour exchange interaction is shown to have a noncollinear ordered ground state with the sublattice magnetization suppressed by quantum fluctuations \cite{Num1, Num2, Num3}. TLAF can be further tuned to QPT into the spin-liquid state by including additional next-nearest \cite{NNN99} or ring exchange interactions \cite{Ring99}, impurities or charge fluctuations \cite{Hubb08}. Another class of systems which show a non-colinear (helimagnetic) order are rare earth elements \cite{Coqblin}.

Previous investigations of frustrated and non-collinear antiferromagnets mainly concentrated on the critical behaviour near the magnetic phase transition. The nonlinear sigma (NL$\sigma$) model \cite{A93} and Landau-Ginzburg-Wilson (LGW) approach \cite{K88,V04} were applied to this problem. These approaches predict critical exponents which are different from those of the standard $O(3)$ universality class. Recent investigations within the non-perturbative renormalization group approach predict, however, a first-order phase transition in three dimensions \cite{NPRG04}. The latter method was also applied to the description of temperature properties near the quantum first-order transition in two-dimensional (2D) frustrated antiferromagnets \cite{F06}. 

The consideration of physical systems requires  a description of a broad temperature- and control parameter range. Even not too close to QPT one can differentiate the high- and low- energy degrees of freedom. While the latter correspond to ``slow'' degrees of freedom and can be described by a continuum model, the former, ``fast'' degrees of freedom, can be absorbed into renormalization of parameters of the model. Thermodynamic properties at low temperatures can be expressed in this way as universal functions of ground state parameters. The corresponding functions describing quantum antiferromagnets can be obtained using the above mentioned NL$\sigma$ model. In the classical case this model corresponds to a continuum limit of the Heisenberg model. For the collinear quantum antiferromagnets the NL$\sigma$ model was first derived by Haldane in the framework of the $1/S$--expansion \cite{H83,A94}. Despite the way of derivation, this model is applicable in both the symmetry broken and the symmetric phases. 

The finite temperature properties may become non-universal in certain cases, e.g., near first order phase transitions, and also for systems described by the quantum field theories at and above their upper critical dimension. The latter theories are non-renormalizable and therefore lead to non-universal properties. Especially interesting example of such a possibility occurs when the theory contains two types of interaction terms, one of which is below and another is above their upper critical dimension. While the former interactions are infrared relevant and produce universal contributions to physical properties, the latter produce non-critical but non-universal contributions. As it is shown in this paper, such a situation is realized in frustrated spin systems, in particular TLAF.
 
The noncollinear magnetic order in frustrated systems is described by so-called $O(3)\otimes O(2)/O(2)$ NL$\sigma$ model \cite{DR89}. This model includes only terms appearing in the large-$S$ limit and is therefore quasiclassical. In comparison to the $O(3)/O(2)$ model for the square lattice this model has different $O(3)\otimes O(2)$ symmetry and a matrix order parameter \cite{A93}. For the description of physical systems the above discussed quasiclassical model should be extended to include all the relevant terms which turn out to be responsible for non-universality in certain parameter range. The construction of the quantum $O(3)\otimes O(2)/O(2)$ NL$\sigma$ model allows one to investigate temperature behaviour of physical properties (e.g., correlation length) above the symmetry broken ground state and near QPT. The implementation of this task, however, calls for the application of non-perturbative methods to deal with the proximity to symmetric phases \cite{Pol75} and critical fluctuations \cite{CSY94}. 

An useful non-perturbative tool of investigating the low-temperature properties of NL$\sigma$ models of spin systems is the $1/N$-expansion, $N$ being the number of the spin components ($N=3$ for the physical case) \cite{Baxter,C85,N88, Z96, Z03, P99}. Contrary to the field-theoretical renormalization group (RG) approach, this method does not suppose universality and renormalizability of the model. For classical case this method introduces a single constraint $\sum_{i}\mathbf{n}_{i}^{2}=\mathcal{N}$ ($\mathbf{n}$ is the fluctuating unit length order parameter, $\mathcal{N}$ is the number of sites) which is exact in the limit $N\rightarrow \infty $ \cite{SpherModelBK,SpherModelS}. In spite of fluctuations of spin length, this approximation appeared to be surprisingly good for description of thermodynamic properties of collinear antiferromagnets \cite{N88,CSY94,IK97,IK98}. Considering $1/N$-corrections allows to obtain in addition a correct description of the critical region. Note that previous considerations of the thermodynamic properties of frustrated antiferromagnets within the $1/N$-expansion in the $\mathbb{CP}^{N-1}$ formulation were constrained to a state with a higher $O(3)\times O(3)$ symmetry \cite{CSS94}. 

The $1/N$-expansion is to some extent complementary to the RG approach. Indeed, the $2+\epsilon$ RG method is perturbative in coupling constant and constrained to the vicinity of two space-time dimensions, but it is non-perturbative in the number of the components. In contrast, the $1/N$-expansion is non-perturbative in the coupling constants and has no restriction on the dimension. Therefore we consider both the methods in our study of frustrated spin systems to have an occasion of comparing their results. Simultaneous using these methods enables one to determine  physical properties in a broad temperature and parameter region.

The content of the paper is as follows. In Sect. 2 we discuss the quasiclassical $O(3)\otimes O(2)/O(2)$ NL$\sigma$ model and its extension to the quantum case. In Sect. 3 the generalization of the quantum NL$\sigma$ model to the $N$-component order parameter and a possibility of performing its $1/N$-expansion are considered. We examine the saddle point $N=\infty$ approximation and consider its stability in Sect. 4, then obtain fluctuation corrections to the Green's functions in Sect. 5. In Sect. 6 we study the renormalizability of the model in the ground state and obtain the RG-flow phase diagram in both the $1/N$-expansion and $2+\varepsilon$ RG approaches. In Sect. 7 we apply the $1/N$-expansion (and the RG approach where it is possible) to calculate correlation length in the renormalized classical and quantum critical regimes. The application of the results to TLAF is considered in Sect. 8. Our results are summarized and discussed in Sect. 9.

\section{Effective action}

As a prototype of frustrated systems, we consider TLAF described by the Heisenberg Hamiltonian 
\begin{equation}
H_{\text{Heis}}=J\sum_{\langle i,j\rangle}\mathbf{S}_{i}\mathbf{S}_{j},  \label{heis}
\end{equation}
where $J>0$ is the antiferromagnetic exchange parameter, $\mathbf{S}_{i}$ are spin operators
on the sites of the triangular lattice, and $\langle i,j\rangle$ denotes the summation over nearest neighbours. Even in the classical case, due to frustration, there are no configurations that would minimize energy of exchange interactions between all nearest neighbours. This results in the noncollinear ground states 
\begin{equation}
\langle \mathbf{S}_{i}\rangle _{0}=S(\mathbf{e}_{1}\cos (\mathbf{Qx_{i}})+\mathbf{e}_{2}\sin (\mathbf{Qx_{i}})),  \label{120str}
\end{equation}
where $\mathbf{Q}=(2\pi /3,-2\pi /\sqrt{3})$ is the wave vector of the AFM structure, $\mathbf{x}_{i}$ is the radius vector of $i$-th site, and $\mathbf{e}_{1},$ $\mathbf{e}_{2}$ are two orthonormal vectors ($\mathbf{e}_{1}^{2}=\mathbf{e}_{2}^{2}=1$, $\mathbf{e}_{1}\cdot\mathbf{e}_{2}=0$) which
determine the plane of spin alignment. An example of such structure is shown in Fig. 1.

\begin{figure}[hpt]
\begin{center}
\includegraphics[scale=1]{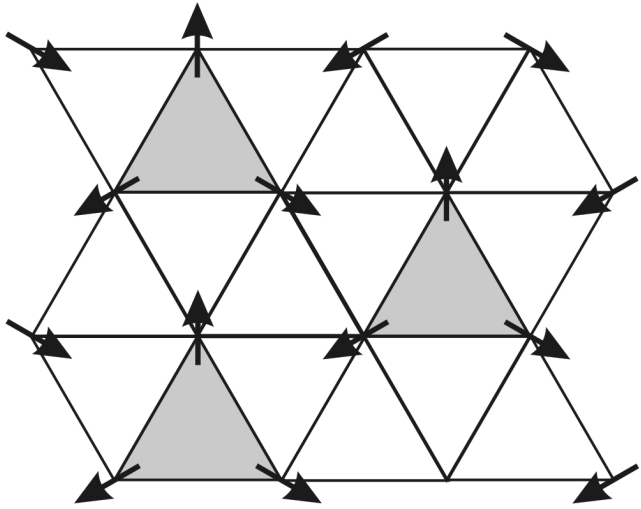}
\end{center}
\caption{The classical ground state magnetic and corresponding plaquette long-range orders}
\end{figure}

Quantum fluctuations lead to the suppression of the sublattice magnetization $\langle \mathbf{S}_{i}\rangle _{0}\rightarrow Z\langle \mathbf{S}_{i}\rangle _{0}$ ($Z<1$, $Z=1$ in the classical limit). For example, in the extremely quantum case $S=1/2$ the result of SWT in the second order in $1/S$ is $Z=0.53$ \cite{C94} (for non-frustrated square lattice with $S=1/2$ the corresponding result is $Z=0.61$ \cite{I92})

The effective low-energy and long-wave-length field theory for the Hamiltonian (\ref{heis}) was derived in Ref. \cite{DR89} in the quasiclassical limit $S\rightarrow \infty $ using the Haldane mapping \cite{H83,A94}. The corresponding action has the form of a NL$\sigma$ model 
\begin{equation}
S_{\text{DR}}=\frac{1}{2c_{\text{in}}g}\int_{0}^{1/T}d\tau \int d^{2}x\left[ |\partial
_{\tau }\mathbf{e}_{1}|^{2}+|\partial _{\tau }\mathbf{e}_{2}|^{2}+|\partial
_{\tau }\mathbf{e}_{3}|^{2}+c_{\mathrm{in}}^{2}(|\nabla \mathbf{e}%
_{1}|^{2}+|\nabla \mathbf{e}_{2}|^{2})\right],  \label{DR}
\end{equation}
where $g=6a/S$ is the coupling constant ($a$ is lattice parameter), orthogonal vectors $\mathbf{e}_{1}$, $\mathbf{e}_{2}$ now acquire space-time coordinate dependence, $\mathbf{e}_{3}=[\mathbf{e}_{1}\times \mathbf{e}_{2}]$, $c_{\text{in}}$ is the in-plane spin-wave velocity. It is assumed that there is some regularization procedure for the wave vectors and frequencies (e.g. using a cutoff parameter $\Lambda$).  

Since quantum effects are not small for physical systems, other terms in the action (\ref{DR}) may be present, besides the renormalization of the spin-wave velocity $c_{\mathrm{in}}$ and the coupling constant $g$. One can obtain corresponding quantum action using symmetry properties of the original Hamiltonian (\ref{heis}) with respect to the $SO(3)$ rotations in the spin space, translations and the time reversal $\tau \rightarrow -\tau $, $\mathbf{e}_{1}\rightarrow -%
\mathbf{e}_{1}$, $\mathbf{e}_{2}\rightarrow -\mathbf{e}_{2}$, as well as the
transformation 
\begin{equation}
\begin{split}
& \mathbf{e}_{1}\rightarrow \mathbf{e}_{1}\cos \varphi +\mathbf{e}_{2}\sin
\varphi \\
& \mathbf{e}_{2}\rightarrow -\mathbf{e}_{1}\sin \varphi +\mathbf{e}_{2}\cos
\varphi,
\end{split}
\label{trfm}
\end{equation}
where $\varphi =2\pi m/3$ ($m$ is an integer number); the latter symmetry follows from the invariance of the model (\ref{heis}) with respect to the translation by the lattice vector. The action, which is invariant under these operations, has the form: 
\begin{eqnarray}
S&=&\frac{1}{2}\int_{0}^{1/T}d\tau \int d^{2}x\left[
p_{1}(|\partial _{\tau }\mathbf{e}_{1}|^{2}+|\partial _{\tau }\mathbf{e}%
_{2}|^{2})+p_{2}|\partial _{\tau }\mathbf{e}_{3}|^{2}+\right.
\label{qcaction} \\
&&+\left. q_{1}(|\nabla \mathbf{e}_{1}|^{2}+|\nabla \mathbf{e}%
_{2}|^{2})+q_{2}|\nabla \mathbf{e}_{3}|^{2}\right],  \notag
\end{eqnarray}
where we have included only the second-order derivative terms. Coefficients $p_{1,2}$ and $q_{1,2}$ are related to bare spin stiffnesses and susceptibilities of the model (\ref{heis}) (see Sect. 3). The action (\ref{qcaction}) neglects the contribution of the Berry phase term, which is expected to be not important for the ordered ground state and low $T$ \cite{S99}. Indeed, singular configurations $\mathbf{e}_{\alpha }$ are separated by an energy gap of order of $J$ from the ground state.  

The effective action (\ref{qcaction}) can be related to the so-called massive $\mathbb{CP}^{M-1}$ model proposed in another description of frustrated systems in terms of the spinon field, connected to the order parameter field $\mathbf{e}_{\alpha}$ by the local two-valued function \cite{CSS94}. However, due to the two-valued character of this mapping, the massive $\mathbb{CP}^{M-1}$ model has to be accomplished by the $\mathbb{Z}_{2}$ lattice gauge field, which provides confinement of spinons \cite{FS79}. In the following we will consider the field theory (\ref{qcaction}) which describes spinons in the confined phase and do not consider the possibility of their deconfinement \cite{ChuStar96}. 

\section{Generalization of the action to the $N$-component order parameter}

\label{Gen_to_N} 
To generalize Eq. (\ref{qcaction}) to the $N$-component case we consider the action
\begin{eqnarray}
S_{\text{gen}} &=&\frac{1}{2}\int_{0}^{1/T}d\tau \int d^{2}x\left[
p_{1}(|\partial _{\tau }\mathbf{e}_{1}|^{2}+|\partial_{\tau }\mathbf{e}_{2}|^{2})+p_{2}(|\partial_{\tau }\mathbf{e}_{3}|^{2}+\cdots +|\partial_{\tau }\mathbf{e}_{N}|^{2})\right.  \notag \\
&&\left. +q_{1}(|\nabla \mathbf{e}_{1}|^{2}+|\nabla \mathbf{e}_{2}|^{2})+q_{2}(|\nabla \mathbf{e}_{3}|^{2}+\cdots +|\nabla \mathbf{e}_{N}|^{2})\right] ,  \label{gen_to_N}
\end{eqnarray}
with $\mathbf{e}_{\alpha }$ ($\alpha=1,\ldots ,N$) being $N$-component vectors forming the orthogonal basis. 
We follow the construction of the $1/N$-expansion in the collinear case \cite{P99, CSY94, S99}, i.e., introduce Lagrange multipliers $\lambda _{\alpha \beta }$ for the constraints $\mathbf{e}_{\alpha }\mathbf{e}_{\beta }=\delta _{\alpha \beta }$ and calculate the Gaussian integral with respect to $\mathbf{e}_{\alpha }$. Since, contrary to the collinear case, $\lambda _{\alpha \beta }$ is a $N\times N$-matrix, even the $N\rightarrow\infty$ limit is non-trivial: it is necessary to sum up an infinite subsequence of the so-called planar diagrams, which is a rather difficult task \cite{C85,P99}.

To formulate an appropriate $1/N$-expansion we consider first the case $q_{2}=0$, $p=p_{1}=p_{2}$ which holds for large enough spin $S$. In this case it is possible to integrate out the vector fields $\mathbf{e}_{3},\ldots \mathbf{e}_{N}$ exactly (note that only $\mathbf{e}_{1}$ and $\mathbf{e}_{2}$ appear in Eq. (\ref{120str}))
and obtain the effective action (see Appendix A): 
\begin{equation}
S_{\text{QEAF}}=\frac{1}{2}\int_{0}^{1/T}d\tau \int d^{2}x\left[ J_{1}^{\mu }(|\partial
_{\mu }\mathbf{e}_{1}|^{2}+|\partial _{\mu }\mathbf{e}_{2}|^{2})-2J_{2}^{\mu
}|(\mathbf{e}_{1},\partial _{\mu }\mathbf{e}_{2})|^{2}\right],   \label{disk}
\end{equation}
where $\mu =\tau ,x,y$, $J^{\tau}_{1}=2p,$ $J^{x}_{1}=J^{y}_{1}=q_{1},$ $J^{\tau}_{2}=p,$ and $J^{x}_{2}=J^{y}_{2}=0.$ The action (\ref{disk}) is justified also for $N=3$, which can be verified by direct substitution $\mathbf{e}_{3}=[\mathbf{e}_{1}\times \mathbf{e}_{2}]$ into Eq. (\ref{DR}); its classical analog was studied in Ref. \cite{A93}. Since (\ref{disk}) contains only two fields, the $1/N$-expansion can be constructed for this action (see Sect.\ref{SPA}). Note that in both the cases $J^{\mu}_{2}<0$ and $J^{\mu}_{2}>0$ the action (\ref{disk}) is stable (in the latter case due to the presence of the nonlinear constraints).

The above-discussed procedure of integration over the vector fields $\mathbf{e}_{3},\ldots, \mathbf{e}_{N}$ cannot be easily generalized to arbitrary $p_{1,2}$ and $q_{1,2}$. In this case we find that the effective action contains not only terms with squares of gradients (as in Eq. (\ref{disk})), but also additional terms with higher gradient powers. We assume that these terms give subleading contribution to physical properties and restrict ourselves to the action (\ref{disk}) with $J_{\alpha }^{x}=J_{\alpha }^{y}$ ($\alpha=1,2$) and arbitrary $J^{\mu}_{\alpha}$ ($\mu=x,\tau$), referred in the following as quantum effective action for frustrated systems (QEAF). This action is a generalization of the quasiclassical action considered by Azaria et al. \cite{A93}. The coupling constants $J_{\mu }^{\alpha}$ are related to the bare spin stiffnesses $\rho^{0}_{\text{in},\text{out}}$ and susceptibilities $\chi^{0}_{\text{in},\text{out}}$ 
\begin{align}
J_{1}^{\tau}&=\chi^{0}_{\mathrm{out}},& J_{1}^{x}&=\rho^{0}_{\mathrm{out}}, \nonumber\\
J_{2}^{\tau}&=\chi^{0}_{\mathrm{out}}-1/2\chi^{0}_{\mathrm{in}},& J_{2}^{x}&=\rho^{0}_{\mathrm{out}}-1/2\rho^{0}_{\mathrm{in}},
\label{J}
\end{align}
where the subscripts `in' and `out' correspond to the in-plane and out-of-plane modes \cite{C94, CSS94}.
QEAF has a form of $O(N)\otimes O(2)/O(N-2)\otimes O(2)_{\text{diag}}\sim O(N)/O(N-2)$ NL$\sigma$ model, $O(2)$-symmetry being defined by the relations (\ref{trfm}) with arbitrary $\varphi$. Hence, we find two types of terms in  QEAF. The first one has the same structure as in the original action (\ref{gen_to_N}). The second (current) term introduces a dynamical quartic interaction additionally to the kinematic one, the latter originating from the nonlinear constraints. This term is related to the square of conserved Noether' current $\mathcal{J}_{12}^{\mu }=(J_{2}^{\mu }-J_{1}^{\mu })(\mathbf{e}_{1},\partial _{\mu }\mathbf{e}_{2})$ for $O(2)$-symmetry.

\section{$1/N$-expansion and solution at $N\rightarrow \infty $}

\label{SPA} To construct the $1/N$-expansion, it is convenient to decouple quartic term $(\mathbf{e}_{1},\partial _{\mu }\mathbf{e}_{2})^{2}$ introducing the Hubbard-Stratonovich field $A_{\mu }$:
\begin{gather}
\mathcal{L}=\frac{N}{2}\mathbf{e}_{\alpha }X^{\alpha \beta }\mathbf{e}%
_{\beta }-\frac{N}{2}\Spur{\lambda}, \\
X_{\alpha \beta }=-(\bar{J}_{1}^{\mu }-\bar{J}_{2}^{\mu })\partial _{\mu
}^{2}\delta _{\alpha \beta }+\bar{J}_{2}^{\mu }[(i\partial _{\mu }-\sigma
^{y}A_{\mu })^{2}]_{\alpha \beta }+\lambda _{\alpha \beta },  \label{X}
\end{gather}%
where $\sigma_{\alpha\beta}^{y}$ is the Pauli matrix, 
\begin{equation}
\bar{J}_{1}^{\mu }=\frac{1}{N}J_{1}^{\mu },\qquad \bar{J}_{2}^{\mu }=\frac{1%
}{N}J_{2}^{\mu },
\end{equation}
and $\lambda_{11},\lambda_{12},\lambda _{22}$ are the Lagrange multipliers for the constraints $\mathbf{e_{1}}\cdot \mathbf{e_{2}}=0$, $\mathbf{e_{1}}^{2}=\mathbf{e_{2}}^{2}=1.$ Integration over $A_{\mu }$
is performed along the real axis for $\bar{J}_{2}^{\mu }>0$ and along the
imaginary axis for $\bar{J}_{2}^{\mu }<0;$ the integration over $\lambda $ is
performed along the imaginary axis. To deal with the symmetry broken phase it is
convenient to represent the field $\mathbf{e}_{\alpha}$ as a sum of a fluctuating field $\mathbf{\tilde{e}}_{\alpha }$ and the uniform static order parameter $\mathbf{m}_{\alpha }$ which is chosen to satisfy $\int dx\mathbf{e}_{\alpha }=\mathbf{m}_{\alpha}\int dx$. Integrating out the field $\mathbf{\tilde{e}}_{\alpha }$ we obtain the following
representation for the partition function: 
\begin{equation}
Z=\int D[\lambda _{\alpha\beta }]\int D[A_{\mu }]\int D[\mathbf{m}_{\alpha
}]\exp {\left( -S_{\text{eff}}[\lambda ,A,\mathbf{m},\mathbf{\tilde{j}},\mathbf{h}%
]\right) },  \label{pf}
\end{equation}
\begin{eqnarray}
\label{Seff}
S_{\text{eff}}[\lambda ,A,\mathbf{m},\mathbf{\tilde{j}},\mathbf{h}] &=&\frac{N}{2}%
\spur{\ln{X}} \\
&&-\frac{N}{2}\int_{0}^{1/T}d\tau \int d^{2}x\left[ \Spur{\lambda(x)}+%
\mathbf{m}_{\alpha }\varphi _{\alpha \beta }\mathbf{m}_{\beta }+\frac{2}{N}%
\mathbf{h}_{\alpha }\mathbf{m}_{\alpha }\right]   \notag \\
&&-\frac{N}{2}(\mathbf{\tilde{j}}_{\alpha }/N-\varphi _{\alpha \delta }%
\mathbf{m}_{\delta })[X^{-1}]_{\alpha \beta }(\mathbf{\tilde{j}}_{\beta
}/N-\varphi _{\beta \gamma }\mathbf{m}_{\gamma }),  \notag
\end{eqnarray}
where we have introduced the sources $\mathbf{\tilde{j}}_{\alpha }$ and $\mathbf{h}_{\alpha }$ of the fields $\mathbf{\tilde{e}}_{\alpha }$ and $\mathbf{m}_{\alpha}$, respectively, and
\begin{equation}
\varphi _{\alpha \beta }=\bar{J}_{2}^{\mu }A_{\mu }^{2}\delta _{\alpha \beta
}-\bar{J}_{2}^{\mu }i\sigma _{\alpha \beta }^{y}\partial _{\mu }[A_{\mu
}]+\lambda _{\alpha \beta }.
\end{equation}
$`\Spur$' denotes the trace over the index $\alpha=1,2$, $`\spur$' includes also an integral over the space-time variables.
 
Calculating the derivatives of Eq. (\ref{pf}) over the source fields we obtain the general expression
for the Green's function $G_{\alpha \beta }^{ij}=\left\langle e_{\alpha}^{i}(x)e_{\beta }^{j}(y)\right\rangle $: 
\begin{equation}
G_{\alpha \beta }^{ij}=\left\langle \frac{1}{N}(X^{-1})_{\alpha \beta
}\delta _{ij}+[(X^{-1}\varphi m^{i})_{\alpha }\otimes (X^{-1}\varphi
m^{j})_{\beta }]+m_{\alpha }^{i}m_{\beta }^{j}\right\rangle,
\label{gfu}
\end{equation}
where $\otimes $ stands for the tensor product with respect to coordinates,
the average is taken with the weight $\exp {(-S_{eff})}$. From the
field-theoretical point of view, this Green's function corresponds to the
propagation of elementary excitations of the symmetry group $O(N)\times O(2)$
with quantum numbers $l=1$, $t=\pm 1$ corresponding to the $O(N)$ and $O(2)$ subgroups, provided that
these symmetries are not broken.

In the $N\rightarrow \infty $ limit the path integral (\ref{pf}) is dominated by the vicinity of saddle points of the functional $S_{\text{eff}}[\lambda ,A,\mathbf{m} ]$, if they exist. The saddle-point equations have the form 
\begin{equation}
\frac{\delta }{\delta \lambda (x)}S_{\text{eff}}=0,\quad \frac{\delta }{\delta A(x)}S_{\text{eff}}=0, \quad 
\frac{\delta }{\delta\mathbf{m}}S_{\text{eff}}=0.
\label{speq}
\end{equation}
The coordinate independent saddle point is $\lambda _{\alpha \beta }=\upsilon ^{2}\delta
_{\alpha \beta }$, $A_{\tau}=A_{x}=A_{y}=0$, $\mathbf{m}_{1}\cdot\mathbf{m}_{2}=0$, $\mathbf{m}_{1}^{2}=\mathbf{m}_{2}^{2}=\sigma _{0}^{2}$, with $\sigma _{0}$ and $\upsilon $ satisfying the equations 
\begin{equation}
\begin{split}
T\sum_{\omega _{n}}\int \frac{d^{2}k}{(2\pi )^{2}}& G_{0}(\mathbf{k},\omega_{n})=1-\sigma _{0}^{2}, \\
& \upsilon \sigma _{0}=0,
\end{split}
\label{mu}
\end{equation}
where
\begin{equation}
G_{0}(\mathbf{k},\omega_{n})=\frac{1}{\bar{J}_{1}^{\tau }\omega _{n}^{2}+\bar{J}_{1}^{x}(
\mathbf{k}^{2}+r_{0}^2)},
\label{g0}
\end{equation}
$\omega _{n}=2\pi nT$ are the bosonic Matsubara frequencies, and $r_{0}^2=\upsilon^2/\bar{J}_{1}^{x}$. It is shown in Appendix B that $\re{S_{\text{eff}}}$ has its smallest value at the saddle point (\ref{mu}), as compared to other possible, in general space-time dependent, saddle points. Therefore the saddle point (\ref{mu}) provides \textit{main} contribution to the partition function at $N\rightarrow \infty $.

The equations (\ref{mu}) coincide with those for the collinear case \cite{CSY94} and were shown to have a solution at $T=0$ with either $\sigma _{0}\neq 0$ (for large enough $J_{1}$), describing the ordered phase, or $\upsilon \neq 0$ (for small enough $J_{1}$) corresponding to the symmetric (paramagnetic) phase. These two states are separated by a second-order quantum phase transition. In agreement with the Mermin-Wagner theorem only symmetric solution is possible at $T>0$. The finite $T$ region above the symmetry broken ground state is denoted as the renormalized classical regime, whereas the region above the quantum critical point is referred to as quantum critical regime \cite{CHN89}.

\section{Fluctuations}

\label{AF} To go beyond the saddle point approximation considered in Sect. \ref{SPA}, we need to take into account fluctuations of the $\lambda$ and $A$ fields. Expanding the action near the saddle point $(\upsilon ^{2},0)$ up to
the second order in fluctuations of these fields, we obtain: 
\begin{eqnarray}
S_{\text{eff}} &=&\frac{N}{4}\int \frac{d^{3}q}{(2\pi )^{3}}\tilde{\Pi}%
(q)(|\alpha _{11}|^{2}+2|\alpha _{12}|^{2}+|\alpha _{22}|^{2})  \notag \\
&&+\frac{N}{2}\int \frac{d^{3}q}{(2\pi )^{3}}\tilde{D}_{\mu \nu }(q)A_{\mu
}(q)A_{\nu }(-q),  \label{efffa}
\end{eqnarray}%
where $\alpha =-i(\lambda -\upsilon ^{2})$, 
\begin{equation}
\tilde{\Pi}(q)=\Pi (q)+2\sigma _{0}^{2}G_{0}(q),\quad \Pi (q)=\int \frac{%
d^{3}p}{(2\pi )^{3}}G_{0}(p)G_{0}(p-q)
\end{equation}
and
\begin{equation}
\begin{split}
& \tilde{D}_{\mu \nu }(q)=D_{\mu \nu }(q)-2\sigma _{0}^{2}\bar{J}_{2}^{\mu }\bar{J}%
_{2}^{\nu }q_{\mu }q_{\nu }G_{0}(q) \\
& D_{\mu \nu }(q)=2\bar{J}_{2}^{\mu }\delta _{\mu \nu }-\bar{J}_{2}^{\mu }%
\bar{J}_{2}^{\nu }\int \frac{d^{3}p}{(2\pi )^{3}}G_{0}(p)G_{0}(p-q)(2p_{\mu
}-q_{\mu })(2p_{\nu }-q_{\nu })
\end{split}
\label{PiBeta}
\end{equation}
are the inverse propagator of the fields $\lambda$ and $A_{\mu}$. We use the notation $p_{\mu }=(\mathbf{p},i\omega _{n})$ and $\int {d^{3}p}/(2\pi )^{3}$ for $T\sum_{n}\int {d^{2}\mathbf{p}}/(2\pi )^{2}$ (and similar for $q$). Hereafter we choose the units $c_{\mathrm{out}}^{2}=1$ and use a regularization of the integrals $\omega ^{2}+\mathbf{p}^{2}<\Lambda ^{2}$. Note that, in contrast to the collinear case, the action (\ref{disk}) is not relativistic invariant, so that the spin-wave velocities will be renormalized by the $1/N$-corrections.

Using the expression for the effective action (\ref{efffa}) we obtain the Green's function 
\begin{equation}
\begin{split}
G_{\alpha \beta }^{ij}(p)=\frac{\delta _{ij}\delta _{\alpha \beta }}{N}\frac{1}{%
G_{0}^{-1}(p)+\bar{J}_{1}\delta r^{2}+\Sigma (p)}  \\
-C_{\alpha \beta }^{(1)ij}(p)-C_{\alpha \beta }^{(2)ij}(p)+\delta
(p)m_{\alpha }^{i}m_{\beta }^{j}. 
\end{split}
\label{gfn}
\end{equation}
The first term in Eq. (\ref{gfn}) is isotropic and non-zero even in the
symmetric phase. The self-energy is given by
\begin{equation}
\Sigma (p)=\Sigma _{1}(p)+\Sigma _{2}(p),  \label{mo}
\end{equation}%
where the contributions 
\begin{equation}
\Sigma _{1}(p)=\frac{3}{N}\int \frac{d^{3}q}{(2\pi )^{3}}\frac{%
G_{0}(p-q)-G_{0}(q)}{\tilde{\Pi}(q)}  \label{S1}
\end{equation}%
and
\begin{equation}
\Sigma _{2}(p)=\frac{1}{N}\int \frac{d^{3}q}{(2\pi )^{3}}\bar{J}_{2}^{\mu }%
\bar{J}_{2}^{\nu }[q_{\mu }q_{\nu }G_{0}(q)-(2p-q)_{\mu }(2p-q)_{\nu
}G_{0}(p-q)]\tilde{D}^{-1}(q)_{\mu \nu }  \label{S2}
\end{equation}
come from the fluctuations of $\lambda_{\alpha\beta}$ and $A_{\mu}$ fields, respectively. The second and third terms
in Eq. (\ref{gfn}) are due to the spontaneous symmetry breaking, 
\begin{gather}
C_{\alpha \beta }^{(1)ij}(p)=G_{0}^{2}(p)m_{\gamma }^{i}m_{\delta
}^{j}\left\langle \alpha _{\alpha \gamma }(p)\alpha _{\beta \delta
}(-p)\right\rangle ,  \label{C1} \\
C_{\alpha \beta }^{(2)ij}(p)=\frac{1}{N}G_{0}^{2}(p)\bar{J}_{2}^{\mu }\bar{J}%
_{2}^{\nu }p_{\mu }p_{\nu }\sigma _{\alpha \gamma }^{y}\sigma _{\beta \delta
}^{y}m_{\gamma }^{i}m_{\delta }^{j}\tilde{D}^{-1}(p)_{\mu \nu }.  \label{C2}
\end{gather}
These terms contribute to the transverse in-plane and longitudinal modes in the leading order in $1/N$ (see \cite{IK97}). In particular, they provide the correct value for $c_{\text{in}}$, which is different from $c_{\text{out}}$. The mass correction $\delta r$ and static order parameter $\mathbf{m}_{\alpha}$ will be calculated in Sect. 7 and Appendix \ref{AppCritInd} in the first order in $1/N$.

For $T=0$ we obtain the propagators of the $\lambda$ and $A$ fields for $q\ll\Lambda$  
\begin{equation}
\begin{split}
& \Pi (q)=\frac{1}{\bar{J}_{1}^{2}}\frac{1}{4\pi q}\arctan \frac{q}{2r_{0}}, \\
& \tilde{D}_{\mu \nu }(q)=\frac{\bar{J}_{2}^{\mu }\bar{J}_{2}^{\nu }}{\bar{J}%
_{1}^{2}}\left\{ M^{\mu }\delta _{\mu \nu }+\left( \delta _{\mu \nu }-\frac{%
q_{\mu }q_{\nu }}{q^{2}}\right) \left[ \pi (q)+2\bar{J}_{1}\sigma _{0}^{2}%
\right] \right\} , \\
& \pi (q)=\frac{q^{2}+4r_{0}^{2}}{8\pi q}\arctan \frac{q}{2r_{0}}{-}\frac{r_{0}}{4\pi},
\end{split}
\label{prop}
\end{equation}
where
\begin{equation}
M^{\mu }=\frac{2\bar{J}_{1}}{\bar{J}_{2}^{\mu }}(\bar{J}_{1}-\bar{J}_{2}^{\mu })+\frac{\Lambda }{3\pi ^{2}}  \label{M}
\end{equation}
is the mass of the field $A_{\mu }$. The last term in Eq. (\ref{M}) originates from our regularization scheme which does not preserve the gauge symmetry for $\bar{J}_{1}=\bar{J}^\mu_{2}$. This term, however, describes the renormalization of coupling constants due to short wave-length fluctuations and therefore should be retained. For $J_{2}^{\mu }\ge0$ we have $M^{\mu }>0$ provided that $\bar{J}_{1}-\bar{J}^\mu_{2}\ge0$, i.e. $\chi^{0}_{\text{in}},\rho^{0}_{\text{in}}\ge0$. The propagators  at $q\lesssim\Lambda$ (when they are needed) can be evaluated numerically.

For $J_{2}^{\mu }\le0$ the field $A_{\mu }=iB_{\mu }$ is purely imaginary and the inverse propagator of the field $B_{\mu }$ equals $-\tilde{D}_{\mu \nu }(q)$. In this case we obtain $M^{\mu }<0$, except for extremely small $J_{1}$ where Eq. (\ref{M}) is not applicable, since we have neglected $r$ in comparison with $\Lambda$; such a small $J_{1}$ are not considered in the following.  For any finite $\Lambda$ and $J_{2}^{\mu}<0$ the propagator $-\tilde{D}_{\mu \nu }(q)$, Eq. (\ref{PiBeta}), is positively defined (this property can be verified by the convolutions of $-\tilde{D}_{\mu \nu }(q)$ with $\delta_{\mu\nu}$ and $q_{\mu}q_{\nu}$), which is in agreement with the stability of the saddle point (\ref{mu}) proven in Appendix B. 
In fact, the propagator $-\tilde{D}_{\mu \nu }(q)$ is a monotonously decreasing function of $q$, having a minimum for  $q=\Lambda$. In the isotropic case $\bar{J}_{2}^{\tau}=\bar{J}_{2}^{x}$ the smallest eigenvalue (``energy gap'') corresponding to $q=\Lambda$ is $2\bar{J}_{1}^2/|\bar{J}_{2}^{x}|+0.015163 \Lambda$. Due to small numerical coefficient this ``energy gap'' can be small enough for $2\bar{J}_{1}^2/|\bar{J}_{2}^{x}|\to 0$.

\section{Renormalization in the symmetry broken state}

\label{RGOP} For practical use of formulae of Sect. \ref{AF}, the dependence of the observable quantities on the ultraviolet cutoff $\Lambda$ should be absorbed into redefined bare parameters of the continuum model. In the present formulation the procedure of renormalization expresses renormalized uniform susceptibilities  $\chi_{\text{in, out}}$, stiffnesses $\rho_{\text{in, out}}$, and the order parameter $|\langle \mathbf{S}\rangle |=Z_{1}|\mathbf{m}_{\alpha}|$ ($Z_{1}$ is some non-universal factor), $\alpha=1, 2$ in terms of the bare parameters of QEAF. This renormalization procedure at $T=0$ is similar to that in the quantum field theory and removes also ultraviolet divergences at finite $T$ \cite{CSY94,IK97}. The infrared divergences in the vicinity of critical points can be removed by additional renormalization.  

To perform renormalization, let us choose the vectors $\mathbf{m}_{i}$ along the axes $i$ ($i=1,2$). Consider the Green's functions $G_{11}^{NN}$ and $G_{11}^{22}$ corresponding to the out- and in-plane modes, respectively. In the leading order in $1/N$ we have (see. Eq. (\ref{gfn})): 
\begin{equation}
G_{11}^{NN}(p)=\frac{1}{N}G_{0}(p)=\frac{\sigma _{0}^{2}}{\chi _{\mathrm{out}}\omega ^{2}+\rho _{\mathrm{out}}\mathbf{p}^{2}}  \label{11}
\end{equation}
with $\chi _{\mathrm{out}}=\sigma _{0}^{2}J_{1}^{\tau }$, $\rho _{\mathrm{out}}=\sigma _{0}^{2}J_{1}^{x}$; $\chi_{\mathrm{out}}=\rho_{\mathrm{out}}$ in our units. Substituting this into Eq. (\ref{mu}) we obtain: 
\begin{equation}
\bar{J}_{1}=\frac{\Lambda }{2\pi ^{2}}+\frac{\rho _{\mathrm{out}}}{N}.
\label{J1}
\end{equation}%
Another relation can be derived from the Green's function 
\begin{equation}
\begin{split}
G_{11}^{22}(p)&=\frac{1}{N\bar{J}_{1}}\frac{1}{p^{2}}\left[ 1-\frac{\bar{J}_{1}\sigma _{0}^{2}}{p/8+2\sigma _{0}^{2}\bar{J}_{1}}\right. \\
&\left. +\bar{J}_{1}\sigma _{0}^{2}\frac{(M^{x}+\tilde{\pi}(p))\omega
^{2}+(M^{\tau }+\tilde{\pi}(p))\mathbf{p}^{2}}{M^{\tau }(M^{x}+\tilde{\pi}
(p))\omega ^{2}+M^{x}(M^{\tau }+\tilde{\pi}(p))\mathbf{p}^{2}}\right]
\end{split}
\label{gin}
\end{equation}
(see Eqs. (\ref{gfn}),(\ref{prop})), where $\tilde{\pi}(p)=p/16+2\rho _{\text{out}}/N$. For small momenta $p\ll
\rho _{\mathrm{out}}$ we find the spin-wave pole of $G_{11}^{22}(p)$: 
\begin{equation}
G_{11}^{22}(p\ll \rho _{out})=\frac{\sigma _{0}^{2}}{\chi _{\mathrm{in}%
}\omega ^{2}+\rho _{\mathrm{in}}\mathbf{p}^{2}},
\end{equation}%
where 
\begin{equation}
\chi _{\mathrm{in}}=\frac{2\chi _{\mathrm{out}}}{1+2\chi _{\mathrm{out}%
}/(M^{\tau }N)},\qquad \rho _{\mathrm{in}}=\frac{2\rho _{\mathrm{out}}}{%
1+2\rho _{\mathrm{out}}/(M^{x}N)}  \label{cr}
\end{equation}%
are the in-plane susceptibility and spin stiffness. From Eq. (\ref{cr}) we see that the masses $M_{x}=2N^{-1}\rho_{\text{out}}/(2\rho_{\text{out}}/\rho_{\text{in}}-1)$ and $M_{\tau}=2N^{-1}\chi_{\text{out}}/(2\chi_{\text{out}}/\chi_{\text{in}}-1)$ of the field $A_{\mu}$ are expressed through the observable quantities. Combining Eqs. (\ref{M}), (\ref{J1}) and (\ref{cr}) we arrive at the result for $J_{2}^{\mu}$ 
\begin{equation}
\bar{J}_{2}^{\mu }=\frac{\bar{J}_{1}^{2}}{C_{\mu }[\bar{J}_{1}-\Lambda
/(2\pi ^{2})]+\Lambda /(3\pi ^{2})},  \label{rgf}
\end{equation}
where $C_{\tau }=[1-\chi _{\mathrm{in}}/(2\chi _{\mathrm{out}})]^{-1}$ and $C_{x}=[1-\rho _{\mathrm{in}}/(2\rho _{\mathrm{out}})]^{-1}$. Supposing $\Lambda$-independence of $\rho_{\text{in,out}}$, $\chi_{\text{in,out}}$ and varying $\Lambda$ we arrive at the renormalization group transformation of the action (\ref{disk}), determined in the limit $N\rightarrow \infty $ by  the formulae (\ref{J1}) and (\ref{rgf}). Note that renormalizations of $J^{\tau }_{\alpha}$ and $J^{x}_{\alpha}$ are independent from each other. The generalization of the RG flow for arbitrary $N$ is considered below.

For $C_{\mu}<0$ (i.e. $J_{2}<0$) the denominator of Eq. (\ref{rgf}) vanishes for $\Lambda=\Lambda^{\mu}_{c}$,
\begin{equation}
\Lambda^{x}_{c}=-\frac{3\pi ^{2}\rho_{\mathrm{out}}}{N}C_{x}=\frac{3\pi^2}{2}\left(|M_x|-\frac{2\rho_{\text{out}}}{N}\right)
\end{equation}
(and similar for $\Lambda^{\tau}_{c}$). In fact the RG flow which begins in the region $J_{2}<0$ cannot be extended to $\Lambda>\Lambda^{\mu}_{c}$ since $J_{2}^{\mu}\to-\infty$ for $\Lambda\to\Lambda^{\mu}_{c}-0$. This corresponds to inapplicability of the continuum model (\ref{disk}) at $J_2<0$  and scales $\Lambda>\Lambda^{\mu}_c$, where one cannot find bare parameters corresponding to given renormalized $\rho_{\text{in,out}}$, $\chi_{\text{in,out}}$.
This peculiarity of the flow arises due to increase of the mass $|M_{\mu}|$ by fluctuations (see Eq. (\ref{M})). 

The restriction on the RG flow is similar to that in the $\phi^{4}$-model in $D\ge4$ and quantum electrodynamics where the values of the renormalized coupling constant (charge) $g_{r}>g_{c}(\Lambda)$ are not physically accessible, $g_{c}(\Lambda)$ vanishing in the limit of infinite cutoff \cite{Weinberg}. For the frustrated systems the inapplicability of the continuum QEAF may correspond to nontrivial change of excitation spectrum similar to that  earlier obtained for the TLAF within the series expansion \cite{SWTPlato1} and renormalized spin-wave theory \cite{SWTPlato2}. Note that the scale $\Lambda^{\mu}_{c}$ can be small enough in comparison with the inverse lattice constant. This is in contrast to the collinear antiferromagnets and above considered case $J_2>0$ where the RG flow can be continued to arbitrary large $\Lambda$.

To generalize the presented analysis  to finite $N$ we consider the perturbative $2+\varepsilon$ RG approach (see Appendix \ref{AppRG}). For simplicity, we restrict ourselves to the RG equations with the space-time isotropy, i.e. $J_2^{x}=J_2^{\tau}$ and use the dimension regularization scheme which conserves gauge symmetry. This regularization neglects the ultraviolet cutoff originating from the lattice. The structure of the flow for $N=3$ is shown  in the coordinates $\alpha=\rho^{0}_{\text{in}}/\rho^{0}_{\text{out}}-1$, $\beta=\Lambda/(2\pi^2\rho^{0}_{\text{out}})$ in Fig. 2. The RG trajectories are characterized by the ratio $M^{\text{RG}}_{x}/\rho_{\text{out}}$ where 
\begin{equation}
M^{\text{RG}}_{x}=\frac{1}{N}\frac{2\rho_{\text{in}}}{|\rho_{\text{in}}/\rho_{\text{out}}-1-\alpha_{c}|^{(N-2)/(N-1)}}
\label{MRG}
\end{equation}
is the generalization of the absolute value of the mass $M_{x}$ of the field $A_{\mu}$ introduced above to finite $N$.
The flow along the trajectories is parametrized by $\Lambda/\rho_{\text{out}}$. The RG flow contains a line of trivial fixed points $\beta=0$ and two non-trivial fixed points. The non-trivial gauge-symmetric fixed point ($\text{O(2) GS}$) has coordinates $\alpha=-1$, $\beta=1/(N-2)$ and it is infrared-unstable in both directions. This point belongs to the line $\alpha=-1$ where the mass $M_{x}$ vanishes and the theory is gauge invariant. The second point $C_{+}$ has coordinates $\alpha=\alpha_c=(N-3)/(N-1)$, $\beta=(N-1)/(N-2)^2$ and it is infrared-unstable in one direction ($\alpha=\alpha_c$) and stable along the other direction (critical line) for arbitrary $N$. Hence, it determines an infrared behaviour near the phase transition, corresponding to the bare parameters on the critical line.  The corresponding critical exponents cannot be determined reliably by the $2+\varepsilon$ RG approach \cite{BZ76}. Moreover recent investigations have shown the absence of stable fixed points in three dimensions at small enough $N$ \cite{NPRG04, Ngo08}. Nevertheless, not too close to the quantum critical point the pseudo-universal behaviour is expected, which is correctly described by the $1/N$-expansion. The corresponding exponents in the first order in $1/N$, $\beta=1/2-6/(\pi^2 N)$ and $\quad \nu=1-16/(\pi^2 N)$ (see Appendix \ref{AppCritInd}) coincide with those for the three-dimensional classical phase transition determined in the framework of Landau-Ginzburg phenomenology \cite{K88}. For $N=3$ the values of the exponents are in reasonable agreement with the pseudo-critical exponents obtained within the non-perturbative RG \cite{NPRG04}.
\begin{figure}[hpt]
\begin{center}
\includegraphics[width=14cm]{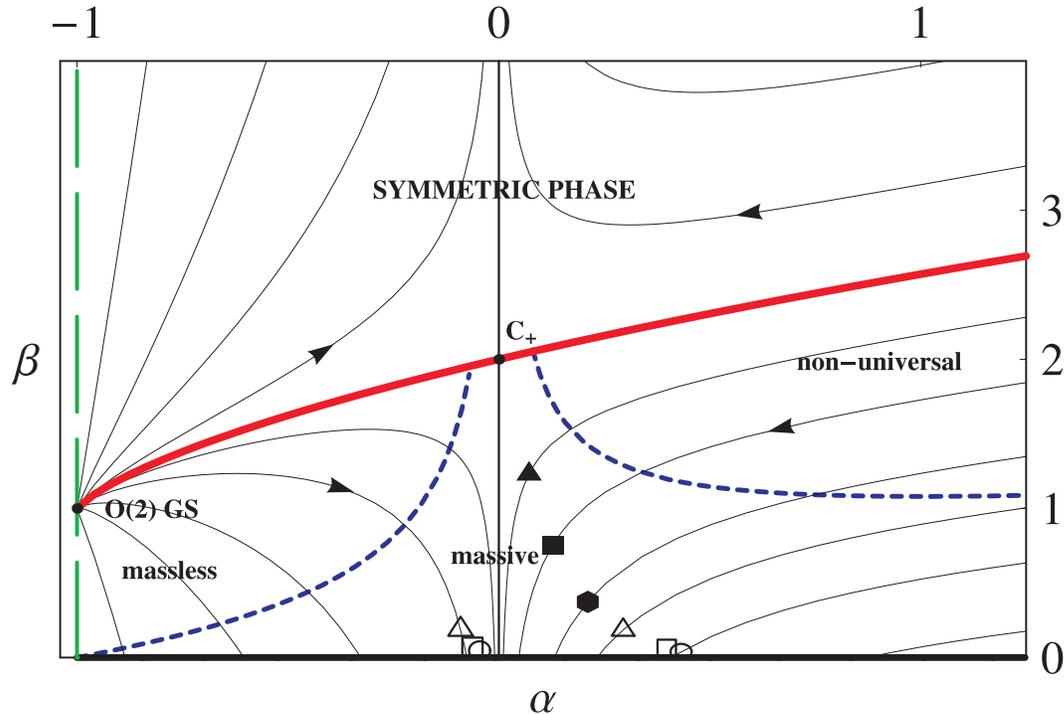}
\end{center}
\caption{The renormalization group flow for $N=3$ and $T=0$ obtained in the perturbative
$2+\varepsilon$ RG approach in the coordinates $\alpha=\rho^{0}_{\text{in}}/\rho^{0}_{\text{out}}-1$, $\beta=\Lambda/(2\pi^2\rho^{0}_{\text{out}})$. The critical (bold) line separates the symmetric (upper part) and symmetry broken (lower part) phases; the fixed point $C_{+}$ controls the (pseudo-) critical behaviour near the phase transition. The mass scale $M^{\text{RG}}_{x}$ vanishes at the long-dashed line; $\text{O(2) GS}$ is the gauge symmetric fixed point. The bold line $\alpha=0$ separates the regions with trivial ($\alpha<0$) and nontrivial ($\alpha>0$) ultraviolet behaviour. ``Massless'' and ``massive'' denote universal regimes with $M^{\text{RG}}_{x}\ll\Lambda/16$ and $M^{\text{RG}}_{x}\gg\Lambda/16$, ``non-universal'' stands for the non-universal regime with $M^{\text{RG}}_{x}\lesssim\Lambda/16$. Short-dashed lines show the crossovers between these regimes. Symbols correspond to the bare parameters of TLAF with different spin values (see Sect.8)}
\end{figure}

Two of four regions of the flow diagram of Fig. 2 correspond to the symmetric, and two to symmetry broken state. 
In the regions with $\alpha<\alpha_c$ the mass $M^{\text{RG}}_{x}$ determines a crossover from the ``massless'' (the part of the region near $\text{O(2) GS}$ fixed point) to ``massive" behaviour of the propagators. On the other hand, at $\alpha>\alpha_c$ where $M^{\text{RG}}_{x}$ cannot be arbitrary small, it determines a change in the RG flow from the universal massive regime at $\Lambda\ll 16 M^{\text{RG}}_{x}$ to the non-universal behaviour at $\Lambda\gtrsim 16 M^{\text{RG}}_{x}$, the latter essentially depends on a cutoff.

Similar to the $N=\infty$ results, the RG flow at $\alpha<\alpha_{c}$ can be continued to arbitrary large $\Lambda$, while, in the right-hand side of the phase diagram ($\alpha>\alpha_{c}$) a critical value of the scale 
\begin{equation}
(\Lambda^{x}_{c})^{\text{RG}}=2\pi \left[\mathcal{F}(+\infty)-\mathcal{F}(\rho_{\text{in}}/\rho_{\text{out}}-1)\right] M_{x}^{\text{RG}},
\label{Lambdac}
\end{equation} 
with the function $\mathcal{F}(\alpha)$ defined in Appendix \ref{AppRG} exists. $(\Lambda^{x}_{c})^{\text{RG}}$ is proportional to the mass $M^{\text{RG}}_{x}$ defined in Eq. (\ref{MRG}), the proportionality factor being non-singular. In this case the RG flow cannot be continued to the region $\Lambda>(\Lambda^{x}_{c})^{\text{RG}}$. As discussed above, for $\Lambda>(\Lambda^{x}_{c})^{\text{RG}}$ there are no bare parameters corresponding to the renormalized $\rho_{\text{in}, \text{out}}$. The value of $\Lambda$ is, however, bounded by $\Lambda\leq\Lambda_{\text{phys}}$ where $\Lambda_{\text{phys}}$ is determined by physical conditions of applicability of the continuum model (for frustrated spin systems $\Lambda_{\text{phys}}\ll a^{-1}$). For $\Lambda_{\text{phys}}<(\Lambda^{x}_{c})^{\text{RG}}$ the action QEAF, Eq. (\ref{disk}) is applicable in the whole admissible range $\Lambda<\Lambda_{\text{phys}}$. In this range the QEAF is renormalizable in the limit $N=\infty$, but already in the first order in $1/N$ this property may be lost, cf. next Section.

\section{Correlation length at finite temperatures}
In this Section we consider the application of the developed $1/N$-expansion to the calculation of correlation length at finite temperatures. 

\subsection{Correlation length in the renormalized classical regime in two dimensions}
In the renormalized classical regime (on the ground-state-ordered side) we expect the exponentially large correlation length for $T\ll \rho_{\text{out}}$, similar to the collinear case \cite{CSY94}. At $N=\infty $ we obtain
\begin{equation}
\xi _{N=\infty }=r_{0}^{-1}=\frac{c_{\text{out}}}{T}\exp \frac{2\pi \rho_{\text{out}}}{NT}.  \label{corr_length}
\end{equation}
Due to the exponential dependence, leading corrections to the correlation length in the first order in $1/N$ come from momenta and frequencies which are much smaller then the scale determined by the temperature. This enables one to take into account only terms with zero Matsubara frequency and use $T/c_{\text{out}}$ as an ultraviolet cutoff for the integrals over momenta. The calculation of the logarithmic corrections follows the same lines as in Chubukov et al. \cite{CSY94}. The correlation length in the first order in $1/N$ is given by
\begin{equation}
\xi ^{-1}=r_{0}+\frac{\delta r^2}{2r_{0}}+\frac{1}{2r_{0}\bar{J}_{1}}\Sigma (\mathbf{k}=ir_{0},\omega
_{n}=0),  \label{ecl}
\end{equation}%
where
\begin{equation}
\delta r^{2}=\frac{1}{\bar{J}_{1}}\frac{R_{A}-R_{\lambda }}{K}, \quad K=\int \frac{d^{3}p}{(2\pi )^{2}}G_{0}^{2}(p),
\label{deltam}
\end{equation}
The quantities $R_{\lambda,A}$ are defined by
\begin{equation}
R_{\lambda }=\int \frac{d^{3}p}{(2\pi )^{3}}G_{0}^{2}(p)\Sigma _{1}(p);\
R_{A}=\int \frac{d^{3}p}{(2\pi )^{3}}G_{0}^{2}(p)\Sigma _{2}(p), \label{R}
\end{equation}
the self-energies $\Sigma_{1,2}$ being defined by Eqs. (\ref{C1}) and (\ref{C2}). In particular, $\Sigma_1$ contains the following logarithmic term 
\begin{equation}
\Sigma _{1}(\mathbf{k}=ir_{0},\omega _{n}=0)\approx -\bar{J}_{1}\frac{3r_{0}^{2}}{2N}
\ln \ln \frac{T}{r_{0}c_{\mathrm{out}}}.
\end{equation}
In contrast, $\Sigma _{2}$ does not contain singularities. Nevertheless both $A_{\mu}$ and $\lambda$ fields contribute logarithmic terms to the mass correction $\delta r^2$. Summing up all the contributions we obtain 
\begin{equation}
\xi_{1/N}^{\text{RC}}\propto \frac{c_{\text{out}}}{T}\left( \frac{NT}{2\pi \rho _{\text{%
out}}}\right) ^{\frac{3}{2N}}\exp \left\{ \left[ 1+\frac{2}{N}+\frac{1}{N}%
\frac{\ln (2\rho _{\text{out}}/\rho _{\text{in}})}{2\rho _{\text{out}}/\rho
_{\text{in}}-1}\right] \frac{2\pi \rho _{\text{out}}}{NT}\right\}, 
\label{xiRC}
\end{equation}
where the last term in the square brackets is due to the $A_{\mu}$ field fluctuations.

In fact, when we turn to smaller values of $N$, both the exponent term and its prefactor in Eq. (\ref{xiRC}) 
are expected to be corrected by higher orders in $1/N$. Instead of direct calculation of these corrections, it is possible to obtain their exact values with the help of RG. For the correlation length we find
\begin{equation}
\xi_{\text{RG}}^{\text{RC}}\propto P(\rho_{\text{in}}/\rho_{\text{out}})\frac{c_{\text{out}}}{T}\left(\frac{NT}{2\pi \rho _{\text{out}}}\right)^{k}\exp \left\{
2\pi\left[\mathcal{F}(\rho_{\text{in}}/\rho_{\text{out}}-1)-\mathcal{F}(\alpha_c)\right]M_{x}^{\text{RG}}/T\right\},
\label{xiRCRG}
\end{equation}
where $k=(6N^3-27N^2+32N-12)/[2(N-2)^{3}(2N-3)]$, $\mathcal{F}(\alpha)$ was defined in Appendix \ref{AppRG}, $P(x)$ is some function (we do not present here its explicit form). We see that the $N$-dependence of the correlation length (\ref{xiRCRG}) is quite complicated, in comparison to the square lattice case \cite{CSY94, CHN89}.

\subsection{Correlation length at finite temperatures above the quantum critical point}

From Eq. (\ref{mu}) we obtain that the inverse correlation length above the quantum critical point $J_1=(J_1)_{c}$ in the zeroth order in $1/N$ is linear in temperature,
\begin{equation}
\xi ^{-1}=r_0=\Theta T,					
\end{equation}
where 
\begin{equation}
\Theta=2\ln\frac{\sqrt{5}+1}{2}=0.962.  
\end{equation}
This behaviour of correlation length is modified by the $1/N$-corrections, as given by Eq. (\ref{ecl}) where $\delta r^2$ now has an additional term coming from the renormalization of the critical value $(J_1)_{c}$ 
\begin{equation}
\delta\tilde{r}^{2}=\frac{R_{A}-R_{A}(T=0)-R_{\lambda}+R_{\lambda }(T=0)}{\bar{J}_{1} K}
\end{equation}
$K=\sqrt{5}/(8\pi T\Theta \bar{J}_{1}^2)$. The general result for the correlation length has the form
\begin{equation}
\xi ^{-1}=\Theta T/c^{\text{r}}_{\text{out}}\left[1+\frac{\gamma(T/\Lambda_{\text{phys}})+W(16 M_{\mu}/\Lambda_{\text{phys}})+U(T/M_{\mu},16 M_{\mu}/\Lambda_{\text{phys}})}{N}\right],
\label{xiQC}
\end{equation}
where $W$ and $U$ are some functions, $\Lambda_{\text{phys}}$ is the upper cutoff scale introduced at the end of the Sect. 6. The $\gamma-$contribution comes from the fluctuations of the $\lambda$ fields ($\Sigma_1$ and $R_{\lambda}$ contributions) and coincides up to the factor $3/2$ with the similar correction to the correlation length of the collinear antiferromagnet \cite{CSY94}, $\gamma(0)=0.3560$ to first order in $1/N$. The function $W$ corresponds to the renormalization of the out-of-plane spin-wave velocity by the zero temperature fluctuations,  $c^{\text{r}}_{\text{out}}=1+N^{-1}W$ in the first order in $1/N$. This renormalization arises because of the absence of the Lorentz invariance leading to the difference of the renormalized and bare out-of-plane spin-wave velocities (the latter is put to unity). The remaining term $U$ comes from the $A$ field fluctuations ($\Sigma_2$ and $R_{A}$ contributions) and leads to the non-trivial temperature dependence of $\xi^{-1}$. One can expect the universal behaviour of the $\xi$ in the regimes $M_{\mu}\ll\Lambda_{\text{phys}}/16$ or $|M_{\mu}|\gg\Lambda_{\text{phys}}/16$, where the sum $W+U$ depends on $T/M_{\mu}$ only. Due to the finite mass of the $A$ field this may lead to a crossover in the temperature dependence of $\xi$. The latter is in contrast to the LGW-model approach where  no current terms leading to additional massive field in the quantum critical region are considered. 

Consider first the ``massless'' regime 
\begin{equation}
M_{\tau,x}\ll\Lambda_{\text{phys}}/16.
\label{mslsReg}
\end{equation}
For frustrated lattice systems this implies  $16 M_{\mu}\ll a^{-1}$ and can be realized only in the left part of the phase diagram
(i.e. for $\rho_{\text{in}}<\rho_{\text{out}}$, $\chi_{\text{in}}<\chi_{\text{out}}$). Condition (\ref{mslsReg}) also corresponds to an abstract field theory where the cutoff parameter should be chosen larger than any other scale. The asymptotics of the integrands allow to conclude that $U$ has a finite limit for $\Lambda_{\text{phys}}\to\infty$. The same assertion holds for $W$. This is in accordance with the renormalizability of the theory at finite temperatures. The temperature dependence of the correlation length for infinite $\Lambda_{\text{phys}}$ was computed numerically and shown in Fig. 3 for $M_{\tau}/M_{x}=1.095$. 
\begin{figure}[hpt]
\begin{center}
\includegraphics[width=14cm]{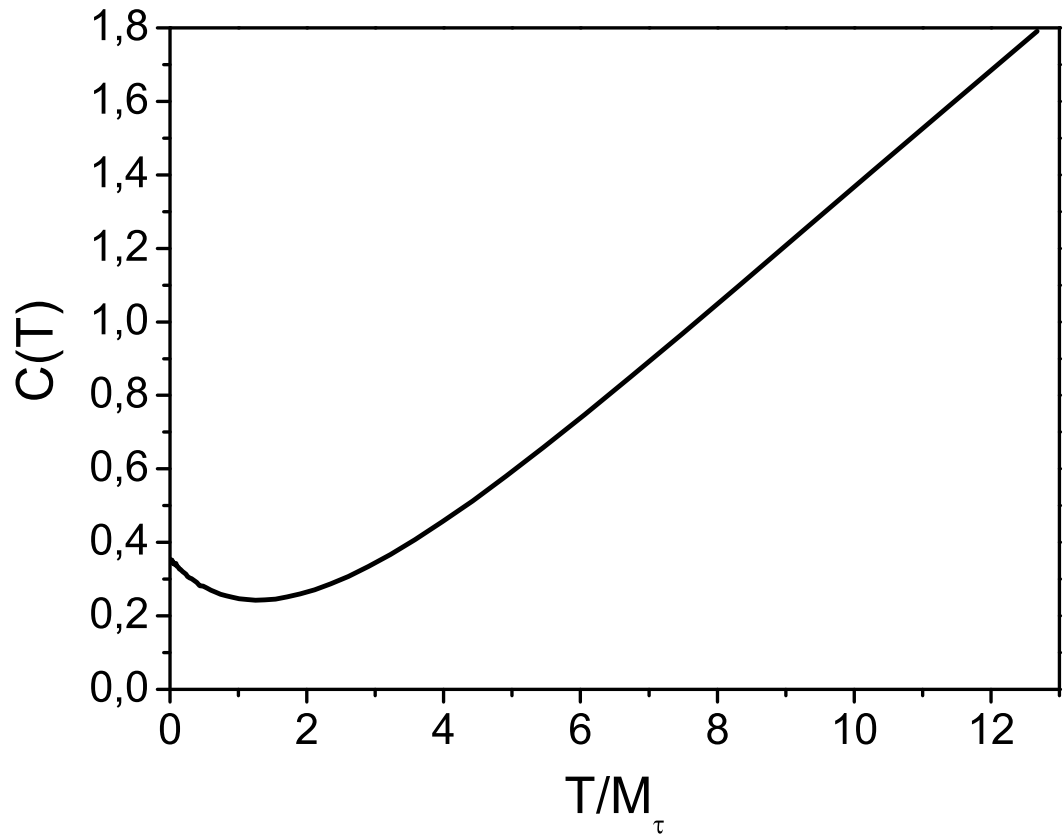}
\end{center}
\caption{The temperature dependence of the coefficient $C=0.3560+W+U$ which determines the correlation length in the quantum critical regime, $\xi ^{-1}=\Theta T/c^{\text{r}}_{\text{out}}\left[1+C/N\right]$ for $M_{\mu}\ll\Lambda_{\text{phys}}/16$ (massless regime) and $M_{\tau}/M_{x}=1.095$} 
\end{figure}
In a broad temperature region $T\gg M_{\tau,x}$ we obtain quadratic behaviour of the inverse correlation length. For $T\approx M_{\tau,x}$ the crossover to the low-temperature regime 
\begin{equation}
\xi ^{-1}(T\to0)=\Theta T/c^{\text{r}}_{\text{out}}\left(1+\frac{0.3560}{N}\right)
\label{xiQC1}
\end{equation}
occurs. Note that in the limit $T\to0$ the contributions $W$ and $U$ cancel each other, and the field $A_{\mu}$ does not contribute to correlation length. 

In the case of large enough $|M_{\tau,x}|\gg\Lambda_{\text{phys}}/16$ (massive regime) the theory is universal at small enough temperatures $T\ll\Lambda_{\text{phys}}/16\ll |M_{\tau,x}|$ in both the right and left part of the phase diagram. Expanding the function $U$ in Eq.(\ref{xiQC}) to first order in $1/M_{\tau,x}$ we find that it contains a non-universal contribution proportional to $\Lambda_{\text{phys}}$,
\begin{equation}
U_{\text{div}}=\frac{7}{15\pi^2}\left( \frac{1}{M_x}-\frac{1}{M_{\tau}}\right) \Lambda_{\text{phys}}.
\end{equation} 
This contribution is, however, cancelled by the similar term in the function $W$ which defines the out-of-plane spin-wave velocity renormalization. The resulting universal expression for the correlation length in the first order in $1/M_{x,\tau}$ is 
\begin{equation}
\xi^{-1}(|M_{x,\tau}|\gg\Lambda_{\text{phys}}/16)=T\Theta/c^{\text{r}}_{\text{out}}\left(1+\frac{0.3560}{N}+\frac{0.0598}{N}\frac{T}{M_{x}}+\frac{0.0473}{N}\frac{T}{M_{\tau}}\right).
\label{CLLargeM}
\end{equation} 
In Eq. (\ref{CLLargeM}) we also find both linear and quadratic terms in $T$. Contrary to the case $|M_{x,\tau}|\ll\Lambda_{\text{phys}}/16$, the $T^2$-term is small in comparison with $T$-term. 

For an intermediate region $|M_{x,\tau}|\sim\Lambda_{\text{phys}}/16$, which corresponds to the crossover region from small $M_{\mu}$ to large $M_{\mu}$ (in the left part of the phase diagram) or to the non-universal regime (in the right part), we obtain the correlation length dependent on $\Lambda_{\text{phys}}$. At low temperatures $T\ll\Lambda_{\text{phys}}/16$ we have  $\xi^{-1}=\Theta T/c^{\text{r}}_{\text{out}}\left[1+(0.3560+\varkappa T)/N\right]$ with $\varkappa$ being some non-universal constant dependent on $M_{\mu}/\Lambda_{\text{phys}}$. At $16 |M_{\mu}|\gg\Lambda_{\text{phys}}$ the value of $\varkappa$ is small and universal, see Eq. (\ref{CLLargeM}). In the right side of the phase diagram for $16|M_{\mu}|\sim\Lambda_{c}\gtrsim\Lambda_{\text{phys}}$ absolute value of $\varkappa$ is slightly larger then in the left part and main contribution to $\xi^{-1}(T)$ is given by the $A_{\mu}$ field fluctuations with momentum and frequencies of order  $\Lambda_{\text{phys}}$. 
 
\section{Application to TLAF}

Now we consider application of the obtained results to TLAF. 
The bare parameters corresponding to TLAF can be obtained using asymptotics $j_1(\Lambda\to0)\to\rho_{\text{out}}/\Lambda$ and $\alpha(\Lambda\to0)\to\rho_{\text{in}}/\rho_{\text{out}}-1$ of Eq. (\ref{AppCRG}) and extending the resulting RG trajectories to larger $\Lambda\leq\Lambda_{\text{phys}}$. The fully renormalized spin stiffnesses and susceptibilities can be obtained from the spin-wave results \cite{C94}
\begin{equation}
\begin{split}
& \chi _{\mathrm{out}}=\frac{2}{9\sqrt{3}J}\left( 1-\frac{0.291}{2S}%
\right) ,\qquad \rho _{\mathrm{out}}=\frac{\sqrt{3}JS^{2}}{4}\left( 1-\frac{%
0.125}{2S}\right) \\
& \chi _{\mathrm{in}}=\frac{2}{9\sqrt{3}J}\left( 1-\frac{0.448}{2S}%
\right) ,\qquad \rho _{\mathrm{in}}=\frac{\sqrt{3}JS^{2}}{2}\left( 1-\frac{%
0.678}{2S}\right),\\
\end{split}
\label{hiSWT}
\end{equation}
rearranged according to the scaling laws $\rho_{\text{in}}=A m^{1/2}$, $\rho_{\text{out}}=A m^{1/2}(1-B m)$ (and similar for spin susceptibilities), where $A$ and $B$ are spin-dependent constants, and $m=1-0.261/S$ is the sublattice magnetization. The bare parameters at $\Lambda=\Lambda_{\text{phys}}$ corresponding to various spin values are marked by different symbols in Fig. 2 (the parameter $\Lambda_{\text{phys}}$ is put to $0.3 a^{-1}$). The open triangular, square and circle denote values of spin stiffnesses (right-hand side of the phase diagram) and susceptibilities (left-hand side) for $S=1/2$, $1$, $3/2$, respectively. Fig. 2 also shows the parameters corresponding to spin values $S<1/2$ which approach QPT at $S=0.261$: the filled triangular, square and hexagon correspond to spins $S=0.267$, $0.29$, and $0.36$. The dependence of both $(\Lambda^{x}_{c})^{\text{RG}}$ and $M^{\text{RG}}_{x}$ on spin is almost linear, $(\Lambda^{x}_{c})^{\text{RG}}\sim 6.36 S-0.34$ and $16 M_{x}^{\text{RG}}\sim 5.07 S - 0.67$; these quantities decrease  when approaching QPT. Near QPT we obtain $(\Lambda^{x}_{c})^{\text{RG}}\sim 16 M^{\text{RG}}_{x}\sim a^{-1}$. To apply the continuum description to this case the inequality $\Lambda_{\text{phys}}\ll a^{-1}\sim 16 M^{\text{RG}}_{x}$ must be satisfied, so that the result (\ref{CLLargeM}) for the correlation length is applicable.

\section{Conclusion}
To describe frustrated spin systems we have considered a generalization of the previously derived quasiclassical $O(3)\times O(2)/O(2)$ model to the quantum case. The corresponding model includes all the relevant terms, some of which  vanish in the classical limit. Additional contributions contain both space and time components of the $O(2)$ current interaction term. In the case of the TLAF they are proportional to the square of the projection of plaquette spin current onto the axis perpendicular to the plaquette. 

To investigate the universality of the finite-temperature properties we have constructed the $1/N$-expansion by introducing a generalized $N>2$ spin-component $O(N)\times O(2)/O(N-2)\times O(2)$ model. Performing the Hubbard-Stratonovich decoupling and introducing Lagrange multipliers for the constraints have led us to the action which contains fluctuating scalar and massive vector fields. The former field is responsible for kinematic interaction which is below its upper critical dimension. The latter field describes dynamic interaction which is above its upper critical dimension. 

We have obtained the flow diagram of Fig. 2 and temperature dependence of the correlation length of the frustrated spin systems within the $1/N$-expansion and $2+\varepsilon$ renormalization group approaches, Eqs. (\ref{xiRCRG}), (\ref{xiQC1}), and (\ref{CLLargeM}). The flow contains massless, massive and non-universal regimes. In the first two regimes physical properties at finite temperature are universal functions of ground-state spin stiffnesses, susceptibilities and order parameter. On the other hand, in the non-universal regime, the dependence on the regularization procedure (e.g. cutoff parameter) cannot be removed by renormalization. When the system is further tuned across the non-universal regime, the continuum model becomes inapplicable at scales $\Lambda>\Lambda_{c}$, where $\Lambda_{c}$ is given by Eq. (\ref{Lambdac}). 

The calculated temperature dependence of the correlation length $\xi$ in the renormalized classical regime is given by  Eqs. (\ref{xiRC}) and (\ref{xiRCRG}). In the quantum critical regime and small enough $M_{\mu}$ (massless regime) 
the correlation length $\xi$ at very low temperatures shows the dependence $\xi\propto T^{-1}$, similar to the collinear case \cite{CSY94}. At larger temperatures this dependence changes to $\xi\propto T^{-2}$. Contrary to the concept of the absence of any scales (except temperature) at the quantum critical point, $M_{\mu}$ leads to a crossover in temperature dependences of physically observable quantities. For large mass $M_{\mu}$ (massive regime) the temperature dependence of $\xi^{-1}$ is almost linear with small quadratic corrections. These  corrections increase and become dependent on the cutoff procedure approaching the non-universal regime of the phase diagram. As it becomes evident in $1/N$-expansion, the (non-)universality is determined by the fluctuations of vector field. 

The effect of the vector field is to some extent similar to that of vortices existing in fact only for $N=3$ \cite{KM84}. At low temperatures the vortices are bounded and therefore not expected to contribute substantially to thermodynamic properties. However, at higher temperatures their effect should be considered. The detailed investigation of the contribution of vortex configurations and their relation to the results obtained in the present paper is the subject of future research. Besides that, real physical systems are not truly two-dimensional, and effects of  interlayer exchange are also to be studied. In the presence of this exchange the long-range order is present at finite $T<T_{\text{Neel}}$, the calculation of the Neel temperature $T_{\text{Neel}}$ should include both spin-wave and vortex contributions.

Acknowledgments. We wish to thank W. Metzner and Yu. Holovatch for discussions. The article is supported by the RFFI grants 07-02-01264a and 1941.2008.2 (Support of scientific schools), and the grant of the Partnergroup of the Max-Planck Society.

\section*{Appendices}
\appendix

\section{Reduction of the action to the $O(N)/O(N-2)$ manifold}

To integrate out rotations of the vector fields $\mathbf{e}_{3},\ldots, 
\mathbf{e}_{N}$ for fixed $\mathbf{e}_{1}$ and $\mathbf{e}_{2}$, it is
convenient to write down the action in the matrix form. Let $i$-th column of
the orthogonal matrix $R(x)$ be constructed from the components of the
vector $\mathbf{e}_{i}(x)$, where $i=1,\ldots, N$. Then Eq. (\ref{gen_to_N}) for $q_{2}=0$, $p=p_{1}=p_{2}$ takes
the form 
\begin{equation}
S_{\text{gen}}=\frac{1}{2}\int dx\left( -p\Spur{\{(R^{T}\partial_{\tau}R)^{2}\}}-q_{1}
\Spur{\{P(R^{T}\nabla R)^{2}\}}\right),  \label{mtrx_form}
\end{equation}%
where $P$ is the projection operator, $P_{11}=P_{22}=1$ and $P_{ij}=0$ in
all the other cases. When $\mathbf{e}_{1}$, $\mathbf{e}_{2}$ are fixed, the
rotation of the vectors $\mathbf{e}_{3},\ldots, \mathbf{e}_{N}$ is described
by the transformation 
\begin{equation}
R(x)\longmapsto R(x)V^{T}(x),  \label{tfmn}
\end{equation}%
where 
\begin{equation}
V(x)=%
\begin{pmatrix}
1_{2} & 0 \\ 
0 & V^{\prime}(x)%
\end{pmatrix},
\label{v1}
\end{equation}%
so that $V^{T}V=1$.

We introduce the following parametrization for $R$. For each matrix $R$ we
consider the matrices $R_{0}$ and $V$ which satisfy the relation $%
R=R_{0}V^{T}$. Thereby we can factorize the group of rotations $O(N)$ in the
factor set $O(N)/O(N-2)$ and the group $O(N-2)$. Then we substitute this
representation of $R$ into (\ref{mtrx_form}). Note that the contributions coming 
from $V$ in the terms with spatial derivatives vanish since they are
cancelled by $P$. As a result we obtain 
\begin{equation}
S_{\text{gen}}[R]=S_{\text{gen}}[R_{0}]+\frac{p}{2}\int dx\left(-\Spur{\{(V^{T}\partial_{\tau}V)^{2}\}}-2\Spur{\{(P^{\prime}R_{0}^{T}\partial_{\tau}R_{0}P^{\prime})^{T}V^{T}\partial_{\tau}V\}}\right),  \label{sw_1}
\end{equation}%
where $P^{\prime}=1-P$ is the projector complementary to $P$. It is important
that the interaction between the fields $R_{0}$ and $V$ is equivalent to the
quadratic interaction between their currents $R_{0}^{T}\partial _{\tau
}R_{0} $ and $V^{T}\partial _{\tau }V$. Besides that, the kinematic
interaction which originates from the integration measure is also possible.
However, as we will show below, all the contributions from the measure of integration
are trivial. Now we replace the variables $V(\tau)$ by $V^{T}\partial_{\tau }V$. Since the space of antisymmetric matrices is flat, the integral over $V^{T}\partial _{\tau }V$ is gaussian. Calculating this integral we arrive at the effective action 
\begin{equation}
S_{\text{\text{QEAF}}}[R_{0}]=S_{\text{gen}}[R_{0}]+\frac{p}{2}\int dx
\Spur{\{(P^{\prime}R_{0}^{T}\partial_{\tau}R_{0}P^{\prime})^{2}\}},  \label{eff}
\end{equation}%
with the order parameter defined on the $O(N)/O(N-2)$ manifold. In Eq. (\ref{eff}) we return to
the $\mathbf{e}_{\alpha}$ representation and obtain the action (\ref{disk}) of the main text with
$J^{\tau}_{1}=2p,$ $J^{x}_{1}=J^{y}_{1}=q_{1},$ $J^{\tau}_{2}=p,$ $J^{x}_{2}=J^{y}_{2}=0.$

Now we show that the measure of integration gives a constant contribution to
the action (\ref{disk}). The transformation of the invariant measure $d\mu (R)$ of the
rotation group has the form 
\begin{equation*}
d\mu (R)=d\mu (R_{0}V^{T})=f(R_{0},V)d\mu (V)d\nu (R_{0}),
\end{equation*}%
where $d\nu $ is the left-invariant measure on the $O(N)/O(N-2)$ manifold
and $f(R_{0},V)$ is the corresponding Jacobian. From the invariance of the
left side of this identity with respect to the transformation $R\longmapsto
gRh$ we have $f(R_{0},V)=f(gR_{0},h^{T}V)$. This means that the Jacobian $f$
is constant and the kinematical interaction between $R_{0}$ and $V$ is
absent: 
\begin{equation*}
d\mu (R)=\text{Const\thinspace }d\mu (V)d\nu (R_{0})
\end{equation*}

It remains to show that the Jacobian of the transformation from $V(\tau )$
to $V^{T}\partial _{\tau }V$ is constant. To this end we write 
\begin{equation*}
\prod_{\tau }d\mu (V(\tau ))=\prod_{\tau }H(V^{T}\partial _{\tau
}V)d(V^{T}\partial _{\tau }V).
\end{equation*}%
The invariance of the left side with respect to the gauge transformation $%
V\longmapsto h(\tau )V$ leads to $H(V^{T}\partial _{\tau }V)=$const.

\section{Global stability of the saddle point}
\label{B}
In this Appendix we show that the saddle point defined by Eq. (\ref{mu}) gives \textit{main} contribution to the partition function (\ref{pf}) for $N\to\infty$. Actually we want to use the following mathematical theorem for the saddle point method \cite{Asymptotic1,Asymptotic2}: if the minimum of the function $\re{\{S_{\text{eff}}\}}$ along the integration contour is achieved \textit{only} at the saddle point $P$, then in the limit $N\rightarrow \infty $ the integral (\ref{pf}) equals to the contribution from this point. We suppose that the system is placed into a finite volume with periodic boundary conditions. In this case we can safely integrate out the gaussian fluctuations of $\sigma_{\alpha}$ in (\ref{pf}). The result of integration equals Eq. (\ref{pf}) apart from $\sigma_{\alpha}=0$. We assume also that the quantity $J^\mu_1-J^\mu_2\ge0$ is nonnegative since it equals to half of the bare in-plane spin stiffness (for $\mu=x$) or susceptibility (for $\mu=\tau$) which are supposed to be positive. 

Consider first the case $J_{2}^{\mu }\ge0$ where the field $A_{\mu}$ is real. We can rewrite the real part of the effective action (\ref{Seff}) as follows:
\begin{equation}
2\re{S_{\text{eff}}[\lambda,A]}=\re{\spur{\ln{\left(1+i K[A]^{-1}L\right)}}}+\spur{\ln{K[A]}},
\label{B_Seff}
\end{equation}
where $K[A]=(\bar{J}_{1}^{\mu }-\bar{J}_{2}^{\mu })(i\partial _{\mu})^{2}+\bar{J}_{2}^{\mu }(i\partial _{\mu }-\sigma ^{y}\mathcal{A}_{\mu})^{2}+\upsilon ^{2}$ is a positively defined operator, $A_{\mu}$ and  $L_{\alpha\beta}(x)=-i(\lambda_{\alpha\beta}(x)-\upsilon^2\delta_{\alpha\beta})$ are arbitrary real functions of space and time variables. The operators $K[A]^{-1}$ and $L$ do not commute in general case. However, since all the eigenvalues $E_{n}$ of the product of the positive and hermitian operators $K[A]^{-1}L$ are real, we obtain for the first term in Eq. (\ref{B_Seff})
\begin{equation}
\re{\spur{\ln{\left(1+i K[A]^{-1}L\right)}}}=\frac{1}{2}\sum_{n}\re{\ln{(1+E^{2}_{n})}}\geqslant0.
\label{B1}
\end{equation}
It can be shown that the sign of equality in this condition stands only for $L=0$. 

The second term in (\ref{B_Seff}) is expressed as a functional integral over complex scalar field $\varphi(x)$
\begin{equation}
\begin{split}
\spur{\ln{K[A]}}&=-2\ln\left(\int D[\varphi^{\ast },\varphi ]\exp \left\{{-\int
dx\varphi ^{\ast }(x)}\left[ {(\bar{J}_{1}^{\mu }-\bar{J}_{2}^{\mu})(i\partial _{\mu })^{2}}\right. \right.\right. \\
&\left. \left. \left. 
{+\bar{J}_{2}^{\mu }(i\partial _{\mu }-\mathcal{A}_{\mu})^{2}+\upsilon ^{2}}\right] {\varphi (x)}^{{}}\right\}  \right). 
\end{split}
\end{equation}
Using transformation of variables $\varphi (x)=\varphi_{r}\exp (i\chi )$ we obtain
\begin{equation}
\spur{\ln{K[A]}}=-2\ln\left(\int\varphi_{r}D[\varphi_{r}]G[A,\varphi_{r}]\exp 
\left\{-\int dx\varphi_{r}(x)\left(-\bar{J}_{1}^{\mu}\partial _{\mu}^{2}+\upsilon ^{2}\right) \varphi_{r}(x)\right\}    \right), 
\label{ff}
\end{equation}
where
\begin{equation}
G[A,\varphi_{r}]=\int D[\chi]\exp \left\{-\int dx\varphi_{r}(x)^2\left[(\bar{J}_{1}^{\mu}-\bar{J}_{2}^{\mu })(\partial_{\mu}\chi )^{2}+\bar{J}_{2}^{\mu}(A_{\mu}-\partial_{\mu}\chi)^{2}\right]\right\} 
\end{equation}
For a fixed $\varphi_{r}$ the maximum of the functional $G[A,\varphi_{r}]$ is reached at $A_{\mu}\equiv0$. Therefore, it follows from Eq. (\ref{ff}), that the minimum of $\spur{\ln{K[A]}}$ is achieved only at $A_{\mu}\equiv0$. Consequently, using Eqs. (\ref{B1}), (\ref{B_Seff}) and aforementioned theorem we conclude the global stability of the saddle point $(\lambda,A)=(\upsilon^{2},0)$ defined in (\ref{mu}) for $J^{\mu}_2>0$. Even if there are some additional saddle points, perhaps with coordinate dependent $A_{\mu}(x)$ and $\lambda(x)$, their contributions to (\ref{pf}) are exponentially small at $N\to\infty$.

For $J^{\mu}_{2}<0$, when the field $A_{\mu}=iB_{\mu}$ is imaginary, we have
\begin{equation}
2\re{S_{\text{eff}}[\lambda,A]}=\re{\spur{\ln{\left(1+i U^{-1}V\right)}}}+\spur{\ln{U}},
\label{B_Seff1}
\end{equation}
where $U=\bar{J}_{1}^{\mu }(i\partial _{\mu})^{2}+|\bar{J}_{2}^{\mu}|B_{\mu}^2+\upsilon ^{2}$, $V=L+|J^{\mu}_{2}|\sigma^{y}(i\partial_{\mu}B_{\mu}+B_{\mu}i\partial_{\mu})$. The above considerations lead us to the stability of the saddle point $(\lambda,A)=(\upsilon^{2},0)$ in this case too.

Note that in the same manner one can prove the global stability of saddle points which arise in the $1/N$-expansions of some nonlinear models, e.g., vector $O(N)$ NL$\sigma$ model and $\mathbb{C}P^{N}$-model. Main requirement
for the applicability of the method is the stability of the action after removing constraints.

\section{Perturbative $2+\varepsilon$ RG approach}
\label{AppRG}
The one loop $2+\varepsilon$ RG equations for QEAF in the case $J_{\alpha}^x=J_{\alpha}^y$ ($\alpha=1,2$) and $T=0$ where obtained in Ref. \cite{A93} 
\begin{equation}
\begin{split}
\frac{dx}{d\ln\Lambda}&=-\varepsilon x+N-\frac{\alpha+5}{2},\\
\frac{d\alpha}{d\ln\Lambda}&=\frac{1+\alpha}{2x}(N-1)\left(\alpha-\alpha_{c}\right),
\end{split}
\label{2pERG}
\end{equation}
where $x=\rho^{0}_{\text{out}}/(K_{2+\epsilon}{\Lambda^{\varepsilon}})\ge0$, $\alpha=\rho^{0}_{\text{in}}/\rho^{0}_{\text{out}}-1\ge-1$, $K_{D}=[2^{D-1}\pi^{D/2}\Gamma(D/2)]^{-1}$, $\alpha_{c}=(N-3)/(N-1)$, bare parameters $\rho^{0}_{\text{in}},\rho^{0}_{\text{out}}$ are related to $J^{x}_{1,2}$ through Eq. (\ref{J}). The solution to Eqs. (\ref{2pERG}) is
\begin{equation}
\begin{split}
&\mathcal{I}=\frac{1}{x}\frac{|\alpha-\alpha_{c}|^{(N-2)/(N-1)}}{1+\alpha}-\frac{2\varepsilon}{n}\mathcal{F}(\alpha)\\
&\Lambda^{\varepsilon}\mathcal{K}=I+\frac{2\varepsilon}{n}\mathcal{F}(\alpha),
\end{split}
\label{AppCRG}
\end{equation}
where $\mathcal{I}$ is the RG invariant ($\mathcal{I}>0$ for the symmetric phase and $\mathcal{I}<0$ for the symmetry broken phase), $\mathcal{K}\ge0$ is an integration constant, 
\begin{align}
\mathcal{F}(\alpha)=\left[\eta(1+\alpha_{c})\right]^{-\frac{N}{N-1}}\Gamma\left(\frac{2N-1}{N-1}\right)&\Gamma\left(\frac{N-2}{N-1}\right)-\notag\\
\left[\eta(1+\alpha)\right]^{-\frac{N}{N-1}}{}_{2}&F_{1}\left(\frac{N}{N-1},\frac{1}{N-1},\frac{2N-1}{N-1},\frac{\alpha_{c}}{\alpha}\right),
\label{gfunc}
\end{align}
where ${}_{2}F_{1}$ is the hypergeometric function, $\eta=1$ for $\alpha>\alpha_{c}$ and $-1$ for $\alpha<\alpha_{c}$. 
A dimensionfull constant $\mathcal{K}$ playing the role of the crossover scale of the RG flow can be expressed through a renormalized parameters of the ordered state $\rho_{\text{in},\text{out}}$ for $\mathcal{I}<0$ using the relations $x=\rho_{\text{out}}/(K_{2+\epsilon}{\Lambda^{\varepsilon}})$, $\alpha=\rho_{\text{in}}/\rho_{\text{out}}-1$ in the limit $\Lambda\to0$. For $\varepsilon=1$ we obtain $\mathcal{K}=1/(\pi^2 N M^{\text{RG}}_{x})$, $M^{\text{RG}}_{x}$ being defined in Eq. (\ref{MRG}). 

An important property of $\mathcal{F}(\alpha)$ is that it has a finite limit for $\alpha\to\infty$. Hence, the RG flow cannot be extended to the region $\Lambda>(\Lambda^{x}_{c})^{\text{RG}}$,
\begin{equation}
(\Lambda^{x}_{c})^{\text{RG}}=\left(\frac{\mathcal{I}+\frac{2\varepsilon}{n}\mathcal{F}(\infty)}{\mathcal{K}}\right)^{1/\varepsilon}.
\label{lmbc}
\end{equation}
Note that this restriction of the RG flow should not be confused with peculiarities of the RG flow originating from the inapplicability of perturbation theory in the symmetric phases. Indeed, the scale $(\Lambda^{x}_{c})^{\text{RG}}$ exists even in symmetry broken phase. For $\mathcal{I}<0$ the scale $(\Lambda^{x}_{c})^{\text{RG}}$ is expressed through the renormalized parameters of the ordered state, as in Eq. (\ref{Lambdac}) 

\section{Critical exponents in dimension $2<D<4$ near QPT}
\label{AppCritInd}
In this Appendix we consider the calculation of critical exponents of QPT in space-time dimensions $2<D<4$ under the assumption that the transition is of second order. 

First we determine the critical exponent for the order parameter $\sigma=|\mathbf{m}_{1}|=|\mathbf{m}_{2}|$ in the symmetry broken phase. The expression for $\sigma$ can be obtained from the identity 
\begin{equation}
\left\langle \mathbf{e}_{\alpha }(x)\mathbf{e}_{\beta }(x)\right\rangle
=\sum_{i=1}^{N}\int \frac{d^{3}p}{(2\pi )^{3}}G_{\alpha \beta
}^{ii}(p)=\delta _{\alpha \beta }
\end{equation}
which follows from the constraint $\mathbf{e}_{\alpha }\mathbf{e}_{\beta
}=\delta _{\alpha \beta }$. Using Eq. (\ref{gfn}) we obtain 
\begin{equation}
\sigma ^{2}=\left[ 1-\int \frac{d^{3}p}{(2\pi )^{3}}G_{0}(p)+R_{\lambda
}-R_{A}\right] (1+F_{\lambda }-F_{A}),  \label{sm}
\end{equation}
where
\begin{gather}
R_{\lambda }=\frac{3}{N}\int \frac{d^{D}q}{(2\pi )^{D}}\frac{I_{1}(q)}{\tilde{\Pi}(q)};\
R_{A}=\frac{1}{N}\int \frac{d^{D}q}{(2\pi )^{D}}%
I_{2}^{\mu \nu }(q)\bar{J}_{2}^{\mu }\bar{J}_{2}^{\nu }\tilde{D}_{\mu \nu
}(p)^{-1}, \label{R0}\\
F_{\lambda }=\frac{1}{N}\int\frac{d^{3}p}{(2\pi )^{3}}\frac{3G_{0}^{2}(p)}{\tilde{\Pi}(p)};\ F_{A}=\frac{1}{N}\int\frac{
d^{3}p}{(2\pi )^{3}}G_{0}^{2}(p)\bar{J}_{2}^{\mu }\bar{J}_{2}^{\nu }p_{\mu }p_{\nu }\tilde{D}^{-1}(p)_{\mu \nu}, \label{F} 
\end{gather}
and 
\begin{equation}
I_{1}(q)=\frac{K_{D}S_{D}(3-D)}{\bar{J}_{1}^{3}}q^{D-6},\qquad I_{2}^{\mu
\nu }(q)=\frac{K_{D}S_{D}}{\bar{J}_{1}^{3}}(q^{2}\delta _{\mu \nu }-q_{\mu
}q_{\nu })q^{D-6},
\end{equation}
\begin{equation}
S_{D}=\frac{\Gamma^{2}(D/2)}{\Gamma(D-1)}\frac{\pi}{\sin{[\pi(D-2)/2}]}.\label{SD}
\end{equation}
In the critical region Eq. (\ref{sm}) takes the form 
\begin{equation}
\sigma ^{2}=\mathrm{Const}\left( 1-\frac{J_{1}^{c}}{J_{1}}\right) \left( 1-%
\frac{1}{N}\frac{3(5-2D)}{(D-2)S_{D}}\ln {\left( 1+\frac{K_{D}S_{D}\Lambda
^{D-2}}{2\sigma ^{2}\bar{J}_{1}}\right) }\right),   \label{sigma}
\end{equation}
where $J_{1}^{c}$ is the critical value of the coupling constant $J_{1}$ which depends on $J_{2}^{\mu }$. It follows from (\ref{sigma}) that the critical exponent $\beta $ for the magnetization in the first order in $1/N$ is given by
\begin{equation}
\beta^{1/N}(D)=\frac{1}{2}+\frac{1}{N}\frac{3}{2S_{D}}\frac{5-2D}{D-2},
\label{beta}
\end{equation}

The critical exponent $\nu $ can be determined from the scaling behaviour of the
spin stiffness
\begin{equation}
\rho _{\text{out}}\propto \left( 1-J_{1}^{c}/J_{1}\right) ^{(D-2)\nu },
\end{equation}
which is determined by Eq. (\ref{11}) with $\sigma_{0}$ being replaced by renormalized order parameter $\sigma$. For $J_{1}-J_{1}^{c}\ll |M^{\mu }|$ we obtain from Eqs. (\ref{gfn}), (\ref{S1}), and (\ref{S2})   
\begin{equation}
G_{11}^{NN}(p\rightarrow 0)^{-1}=\text{Const}^{\prime }J_{1}p^{2}\left( 1-%
\frac{3}{N}\frac{4-d}{d(d-2)S_{D}}\ln {\left( 1+\frac{K_{D}S_{D}\Lambda
^{D-2}}{2\sigma ^{2}\bar{J}_{1}}\right) }\right).
\end{equation}
As a result we obtain
\begin{equation}
\nu^{1/N}(D)=\frac{1}{D-2}\left(1-\frac{6(D-1)}{N S_{D} D}\right).
\label{nu}
\end{equation}

\end{document}